\documentclass[11pt]{article}
\usepackage{moriond,epsfig}

\def\Journal#1#2#3#4{{#1} {\bf #2}, #3 (#4)}

\def\NIMA{{\em Nucl. Instrum. Methods} A}

\def\PLB{{\em Phys. Lett.}  B}
\def\PRL{\em Phys. Rev. Lett.}
\def\PRD{{\em Phys. Rev.} D}
\def\ZPC{{\em Z. Phys.} C}
\def\EPJC{{\em Eur. Phys. J.} C}

\def\ra{\rightarrow}

\def\be{\begin{equation}}
\def\ee{\end{equation}}
\def\bea{\begin{eqnarray}}
\def\eea{\end{eqnarray}}

\newcommand{\kskl}{K_S^0 K_L^0}
\newcommand{\ks}{K_S^0}

\newcommand{\BR}{{\cal B}}

\newcommand{\psp}{\psi^\prime}
\newcommand{\jpsi}{J/\psi}

\newcommand{\chico}{\chi_{c1}}
\newcommand{\chict}{\chi_{c2}}

\newcommand{\EE}{e^+e^-}
\newcommand{\MM}{\mu^+\mu^-}
\newcommand{\ggee}{\gamma\gamma e^+e^-}
\newcommand{\gguu}{\gamma\gamma \mu^+\mu^-}
\newcommand{\Jpi}{\pi^0 J/\psi}
\newcommand{\Jeta}{\eta J/\psi}

\newcommand{\gx}{\gamma\chi_{c1,c2}}

\newcommand{\piz}{\pi^0}
\newcommand{\pp}{\pi^+\pi^-}
\newcommand{\kk}{K^+K^-}
\newcommand{\ppb}{p\overline{p}}
\newcommand{\aab}{\Lambda\overline{\Lambda}}
\newcommand{\ksks}{K^0_S K^0_S}

\newcommand{\jpsito}{J/\psi \rightarrow }

\newcommand{\pspto}{\psp \rightarrow }
\newcommand{\chicJto}{\chi_{cJ} \rightarrow }

\newcommand{\bfg}{\begin{figure}}
\newcommand{\efg}{\end{figure}}
\newcommand{\bitm}{\begin{itemize}}
\newcommand{\eitm}{\end{itemize}}
\newcommand{\bnum}{\begin{enumerate}}
\newcommand{\enum}{\end{enumerate}}
\newcommand{\btbl}{\begin{table}}
\newcommand{\etbl}{\end{table}}
\newcommand{\btbu}{\begin{tabular}}
\newcommand{\etbu}{\end{tabular}}

\begin{document}
\vspace*{4cm}
\title{Recent BES results on charmonium decays}

\author{C.~Z.~Yuan (For the BES Collaboration)}

\address{Institute of High Energy Physics, Chinese Academy of Sciences,\\
Beijing 100039, China}

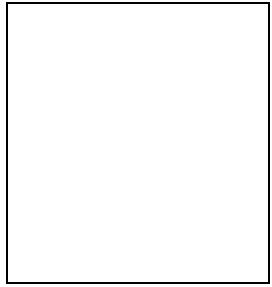
\maketitle\abstracts{Recent results on charmonia decays at
BES/BEPC are reported, including the observation of $\pspto
\kskl$, $\pspto Vector\, \, Tensor$ for the measurement of the
relative phase between the strong and electromagnetic decays of
$\psp$ and a test of the pQCD ``12\% rule'' between $\psp$ and
$\jpsi$ decays; the study of $\pspto \gamma \gamma \jpsi$ for the
determination of $\pspto \piz \jpsi$, $\eta\jpsi$, $\gamma \chico$
and $\gamma \chict$ decay branching fractions; the test of the
color-octet mechanism via $\chicJto \ppb$ and $\chicJto \aab$; and
a search for the CP violating process $\psp$ and $\jpsito \ksks$.
}

\section{BES experiment and the data samples}

The data samples used for the analyses are taken with the Beijing
Spectrometer (BESII) detector~\cite{bes,bes2} at the Beijing
Electron-Positron Collider (BEPC) storage ring at a center-of-mass
energies corresponding to $M_{\psp}$ and $M_{\jpsi}$. The data
samples contain $(14 \pm 0.6)\times 10^6$ $\psp$ events and $(57.7
\pm 2.7)\times 10^6$ $\jpsi$ events, as determined from inclusive
hadronic decays.

\section{Observation of $\pspto \kskl$}

It has been determined that for many two-body exclusive $\jpsi$
decays~\cite{suzuki,jphase,wymphase} the relative phases between
the three-gluon and the one-photon annihilation amplitudes are
near $90^\circ$.  For $\psp$ decays, the available information
about the phase is much more limited because there are fewer
experimental measurements. The analysis of $\pspto Vector \,\,
Pseudoscalar$ (VP) decays shows that the phase could be the same
as observed in $\jpsi$ decays~\cite{wymphase}, but it could not
rule out the possibility that the phase is near $180^\circ$ as
suggested in Ref.~\cite{suzuki} due to the big uncertainties in
the experimental data. A measurement of the relative phase in
$\pspto Pseudoscalar\,\, Pseudoscalar$ (PP) is suggested in
Ref.~\cite{phase_pp} by searching for $\pspto \kskl$.

\begin{figure}[htbp]
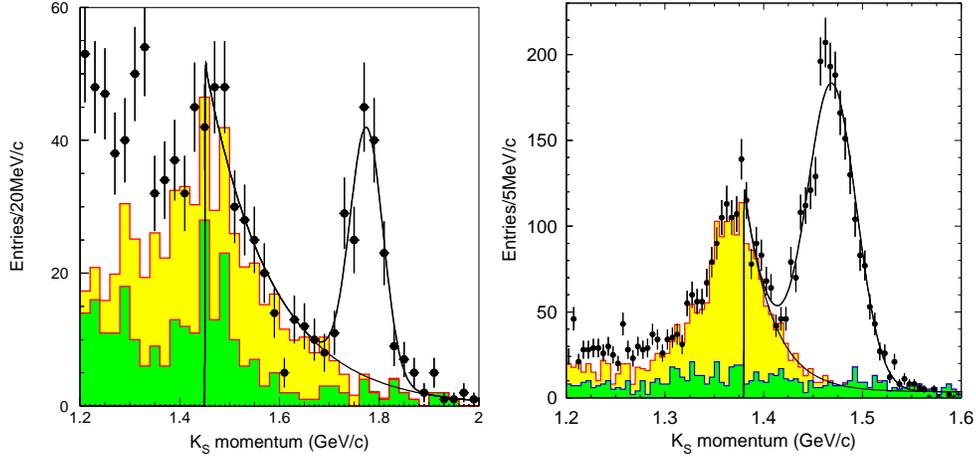

\centerline{\hbox{ \psfig{file=pksdtfit_prl_new.epsi,height=6cm}}
\hbox{ \psfig{file=pks_fit_j_prd.epsi,height=6cm}}} \caption{The
$K_S$ momentum distribution for data at $\psp$ (left) and $\jpsi$
(right). The dots with error bars are data and the curves are the
best fit of the data. The dark shaded histogram is from $K_S$ mass
side band events, and the light shaded histogram is from the Monte
Carlo simulated backgrounds. } \label{kskl}
\end{figure}

BESII searches for $\pspto \kskl$ by reconstructing the monochroic
$\ks$ in the 14~M $\psp$ data sample~\cite{bes2kskl}. The signal,
as shown in Fig.~\ref{kskl}, is very significant (about
13$\sigma$), and the branching fraction is measured to be \(
\BR(\pspto \kskl) = (5.24\pm 0.47 \pm 0.48)\times 10^{-5}\). This
branching fraction, together with branching fractions of $\pspto
\pp$ and $\pspto \kk$, are used to extract the relative phase
between the three-gluon and the one-photon annihilation amplitudes
of the $\psp$ decays to pseudoscalar meson pairs. It is found that
a relative phase of $(-82\pm 29)^{\circ}$ or $(+121\pm
27)^{\circ}$ can explain the experimental results~\cite{phase_pp}.

A similar analysis of the $\jpsi$ data sample yields an improved
measurement of the $\jpsito \kskl$ (see Fig.~\ref{kskl}) branching
fraction~\cite{bes2ksklj}: \( \BR(\jpsito \kskl) = (1.82\pm 0.04
\pm 0.13)\times 10^{-4}\), which is more than $4\sigma$ larger
than the world average~\cite{pdg}. Comparing with the
corresponding branching fraction for $\pspto \kskl$ , one gets
      \( Q_h = \frac{\BR(\pspto \kskl)}{\BR(\jpsito \kskl)}
                   = (28.8\pm 3.7)\% \).
This result indicates that $\psp$ decays is enhanced by more than
4$\sigma$ relative to the ``12\% rule'' expected from
pQCD~\cite{beswangwf}, while for almost all other channels where
the deviations from the ``12\% rule'' are observed, $\psp$ decays
are suppressed.

The violation of the ``12\% rule'' in $\kskl$ mode is explained in
Ref.~\cite{wmykskl} in the $S$- and $D$-wave mixing model of the
$\psp$ state. In this scenario, the $\psi(3770)$, also an $S$- and
$D$-wave mixed charmonium state will have a decay branching
fraction to $\kskl$ between $(0.12\pm 0.07)\times 10^{-5}$ and
$(3.8\pm 1.1)\times 10^{-5}$. This need to be tested with the
large $\psi(3770)$ data samples at CLEOc and BESIII.

\section{Observation of $\pspto Vector \,\, Tensor$}

Four $Vector \,\, Tensor$ (VT) decay channels $\pspto \omega
f_{2}(1270) \rightarrow \pi^+\pi^-\pi^+\pi^-\pi^0$, $\rho
a_2(1320) \rightarrow \pi^+\pi^-\pi^+\pi^-\pi^0$,
$K^*(892)^0\overline{K}^*_2(1430)^0+c.c. \ra \pi^+\pi^-K^+K^-$ and
$\phi f_2^{\prime}(1525) \ra K^+K^-K^+K^-$ are investigated to
test the pQCD ``$12\%$ rule''~\cite{beswangwf}. Previous BESI
results~\cite{BESVT} on these channels reveal that these VT decay
modes are suppressed compared to the perturbative QCD prediction.
However, the measurements, using about $4\times 10^6$ $\psp$
events, determined only upper limits or branching fractions with
large errors. These analyses are updated with $14\times 10^6$
$\psp$ events, and signals of all these four channels are
observed~\cite{bes2vt}.  The statistical significance for all four
channels are larger than $3 \sigma$; those for $\omega f_2(1270)$
and $K^*(892)^0\overline{K}^*(1430)^0+c.c.$ are larger than $5
\sigma$. Table~\ref{BESII_VT} summarizes the results of the four
branching fraction measurements, as well as the corresponding
branching fractions of $J/\psi$ decays, and the ratios of the
$\psp$ to $J/\psi$ branching fractions. All four VT decay modes
are suppressed by a factor of 3 to 5 compared with the pQCD
expectation.

\begin{table}
\caption{\label{BESII_VT}
  Branching fractions measured for
  $\pspto Vector\,\, Tensor$. Results for corresponding
  $J/\psi$ branching fractions are also given as well as the
  ratio $Q_X=\frac{\BR(\pspto X)}{\BR(\jpsito X)}$. }
\begin{tabular}{|c|c|c|c|c|c|}  \hline \hline
X &$N^{obs}$ & $\epsilon(\%)$ & $B(\psp\ra X)(\times
10^{-4})$ & $B(J/\psi\ra X)(\times 10^{-3})$ & $Q_X(\%)$ \\
\hline $\omega f_{2}$  & $62\pm12$  & $4.25\pm0.10$  &
$2.05\pm0.41\pm0.38$
 & $4.3\pm0.6$ & $4.8\pm 1.5$ \\
$\rho a_2$              & $112\pm31$ & $6.42\pm0.06$
& $2.55\pm0.73\pm0.47$ & $10.9\pm2.2$ & $2.3\pm1.1$ \\
$K^*\overline{K}^*_2$   & $93\pm16$ & $16.2\pm0.2$
& $1.86\pm0.32\pm0.43$ & $6.7\pm2.6$  &  $2.8\pm1.3$ \\
$\phi f_2^{\prime}$     & $19.7\pm5.6$  & $14.8\pm0.2$ &
$0.44\pm0.12\pm0.11$
 & $1.23\pm 0.21$ & $3.6\pm1.5$ \\ \hline \hline
\end{tabular}
\end{table}

\section{Analysis of $\pspto \gamma \gamma \jpsi$}

$\pspto\pi^0\jpsi$, $\eta\jpsi$ and $\gamma\chi_{c1,2}$ decay
branching ratios are determined by measuring $\gamma \gamma
\jpsi$, $\jpsito \EE$ or $\MM$ final states~\cite{bes2ggpsi}. The
results are shown in Table~\ref{results}.

\begin{table}[htbp]
\caption{\label{results} Results on $\pspto \gamma \gamma \jpsi$
analysis. }
\begin{center}
{\footnotesize{
\begin{tabular}{c|cc|cc}              \hline        \hline
Channel&\multicolumn{2}{c}{$\pi^0\jpsi$}&\multicolumn{2}{c}{$\eta\jpsi$}\\\hline
Final state&$\ggee$&$\gguu$&$\ggee$&$\gguu$\\\hline
 BR(\%)&$0.139\pm 0.020\pm 0.012$&$0.147\pm 0.019\pm 0.013$&$ 2.91\pm
0.12\pm 0.21$&$3.06\pm 0.14\pm 0.25 $\\\hline Combine BR
(\%)&\multicolumn{2}{|c|}{$0.143\pm 0.014\pm 0.012$}
&\multicolumn{2}{|c}{$2.98\pm 0.09\pm 0.23$}\\
\hline\hline
Channel&\multicolumn{2}{c}{$\gamma\chi_{c1}$}&\multicolumn{2}{c}
{$\gamma\chi_{c2}$}\\\hline Final
state&$\ggee$&$\gguu$&$\ggee$&$\gguu$\\\hline
 BR (\%)&$8.73\pm
0.21\pm 1.00$&$9.11\pm 0.24\pm 1.12$&
$7.90\pm 0.26\pm 0.88$&$8.12\pm 0.23\pm 0.99$\\
Combine BR (\%)&\multicolumn{2}{|c|}{$8.90\pm 0.16\pm
1.05$}&\multicolumn{2}{|c}{$8.02\pm 0.17\pm 0.94$}\\
\hline\hline
\end{tabular}
}}
\end{center}
\end{table}

The BES $\BR(\pspto\Jpi)$ measurement has improved precision by
more than a factor of two compared with other experiments, and the
BES $\pspto\Jeta$ branching fraction is the most accurate single
measurement.  The $\BR(\pspto\Jpi)$ agrees better with the Mark-II
result~\cite{MK2} than with the Crystal Ball result~\cite{CB},
while $B(\pspto\gx)$ agrees well with the Crystal Ball
results~\cite{CB}. The measurements are used to test various
models in calculating the $\psp$ decays rates~\cite{models}.

\section{Test of COM in P-wave charmonium Baryonic decays}

Hadronic decay rates of P-wave quarkonium states provide good
tests of QCD. The decays $\chicJto \ppb$ have been calculated
using different models~\cite{besi5}, and recently, the decay
branching fractions of $\chicJto$ baryon and anti-baryon pairs
were calculated including the contribution of the color-octet fock
states~\cite{wong}.  Using the $\chicJto \ppb$ branching fractions
as input to determine the matrix element, the partial widths of
$\chicJto \aab$ are predicted to be about half of those of
$\chicJto \ppb$, for $J=1$ and $2$.  As shown in Table~\ref{br},
the measurements of $\chicJto \aab$~\cite{aa} together with the
branching fractions of $\chicJto \ppb$~\cite{bes2ppb} from the
same data sample, indicate that $\chicJto \aab$ is enhanced
relative to $\chicJto \ppb$, as compared with the color-octet
mechanism (COM) calculation~\cite{wong}.

\begin{table}[htbp]
\caption{Branching fractions of $\chicJto \aab$ and $\chicJto
\ppb$, and \( R_{\cal B} = \BR(\chicJto \aab)/\BR(\chicJto
\ppb).\)}
\begin{center}
\begin{tabular}{c|ccc}\hline\hline
$\BR(\chicJto \aab)$ ($10^{-5}$) &$47^{+13}_{-12}\pm10$
                                &$26^{+10}_{-9}\pm 6$
                                &$33^{+15}_{-13}\pm 7$  \\
$\BR(\chicJto \ppb)$ ($10^{-5}$)&$27.1^{+4.3}_{-3.9}\pm4.7$
                                &$5.7^{+1.7}_{-1.5}\pm0.9$
                                &$6.5^{+2.4}_{-2.1}\pm1.0$  \\
$R_{\cal B}$&$1.73\pm0.63$&$4.6\pm2.3$&$5.1\pm3.1$\\
\hline\hline
\end{tabular}
\end{center}
\label{br}
\end{table}

\section{Search for $\psp$ and $\jpsito \ksks$}

The CP violating processes $\jpsito \ksks$ and $\pspto \ksks$ are
searched for using the $\jpsi$ and $\psp$
samples~\cite{bes2ksks}. One candidate in each case is observed,
in agreement with the expected background level. The upper limits
on the branching ratios are determined to be \( \BR(\jpsito \ksks)
<1.0\times 10^{-6} \) and \( \BR(\pspto \ksks) <4.6\times 10^{-6}
\) at the 95\% C. L. The former is much more stringent than the
previous Mark-III measurement~\cite{mk3ksks}, and the latter is
the first search for this channel in $\psp$ decays. The current
bounds on the production rates are still far beyond the
sensitivity needed for testing the EPR paradox~\cite{EPR}, and
even farther for CP violation~\cite{roo}.

\section*{Acknowledgments}

Thanks my colleagues of BES collaboration who did the good work
which are reported here, and thanks Profs. J.~Tran Thanh Van and
E.~Aug\'e for the successful organization of the Recontres.


\section*{References}

\begin{thebibliography}{99}

\bibitem{bes} J.~Z.~Bai. {\em et al.} (BES Collab.), \Journal\NIMA{344}{319}{1994}.

\bibitem{bes2} J.~Z.~Bai. {\em et al.} (BES Collab.),\Journal\NIMA{458}{627}{2001}.

\bibitem{suzuki} M.~Suzuki, \Journal\PRD{63}{054021}{2001};
                 J.~L.~Rosner, {\em ibid.} {\bf 60}, 074029 (1999).

\bibitem{jphase} J.~Jousset {\em et al.},
\Journal\PRD{41},{1389}{1990};
 D.~Coffman {\em et al.}, {\em ibid.} {\bf 38}, {2695} {(1988)};
 M.~Suzuki, {\em ibid.} {\bf 60}, {051501}, {(1999)};
 L.~K\"{o}pke and N.~Wermes,
                 {\em Phys. Rep.} {\bf 174}, {67} {(1989)};
 R.~Baldini {\em et al.}, \Journal\PLB{444}{111}{1998}.

\bibitem{wymphase} P.~Wang, C.~Z.~Yuan and X.~H.~Mo,
                  \Journal\PRD{69}{057502}{2004}.

\bibitem{phase_pp} C.~Z.~Yuan, P.~Wang and X.~H.~Mo,
                   \Journal\PLB{567}{73}{2003}.

\bibitem{bes2kskl} J.~Z.~Bai. {\em et al.} (BES Collab.),
              \Journal\PRL{92}{052001}{2004}.

\bibitem{bes2ksklj} J.~Z.~Bai. {\em et al.} (BES Collab.),
              \Journal\PRD{69}{012003}{2004}.

\bibitem{pdg} K.~Hagiwara {\em et al.} (Particle Data Group),
              \Journal\PRD{66}{010001}{2002}.

\bibitem{beswangwf} J.~Z.~Bai. {\em et al.} (BES Collab.),
             \Journal\PRD{67}{052002}{2003}.

\bibitem{wmykskl}P.~Wang, X.~H.~Mo and C.~Z.~Yuan, hep-ph/0402227.

\bibitem{BESVT} J. Z. Bai {\em et al.} (BES Collab.),
      \Journal\PRL{81}{5080}{1998} and \Journal\PRD{67}{052002}{2003}.

\bibitem{bes2vt} J. Z. Bai {\em et al.} (BES Collab.), \Journal\PRD{69}{072001}{2004}.

\bibitem{bes2ggpsi} J. Z. Bai {\em et al.} (BES Collab.), hep-ex/0403023.

\bibitem{MK2} T. Himel {\em et al.}, \Journal\PRL{44}{920}{1980}.

\bibitem{CB} M. J. Oreglia {\em et al.}, \Journal\PRL{45}{959}{1980};
             J. Gaiser {\em et al.}, \Journal\PRD{34}{711}{1986}.

\bibitem{models} G. A. Miller et al., {\em Phys. Rep.} {\bf 194}, {1}
{(1990)}; R. casalbuoni {\em et al.},
\Journal\PLB{309}{163}{1993}; Y. P. Kuang and T. M. Yan,
\Journal\PRD{24}{2874}{1981}; Y. P. Kuang {\em et al.}, {\em
ibid.} {\bf 37}, 1210 (1988).

\bibitem{besi5} M.~Anselmino, R.~Cancelliere, and F.~Murgia,
                 \Journal\PRD{46}{5049}{1992};
                 M.~Anselmino, F.~Caruso, and S.~Forte,
                 {\em ibid.} {\bf 44}, 1438 (1991);
               M.~Anselmino and F.~Murgia, \Journal\ZPC{58}{429}{1993}.

\bibitem{wong} S.~M.~H.~Wong, \Journal\EPJC{14}{643}{2000}.

\bibitem{aa} J.~Z.~Bai. {\em et al.} (BES Collab.),
              \Journal\PRD{67}{112001}{2003}.

\bibitem{bes2ppb} J.~Z.~Bai. {\em et al.} (BES Collab.),
              hep-ex/0401011.

\bibitem{bes2ksks} J.~Z.~Bai. {\em et al.} (BES Collab.),
              \Journal\PLB{589}{7}{2004}.

\bibitem{mk3ksks} R.~M.~Baltrusaitis {\em et al.},
                \Journal\PRD{32}{566}{1985}.

\bibitem{EPR} A.~Einstein, B.~Podolsky and N.~Rosen,
               {\em Phys. Rev.} {\bf 47}, 777 (1935).

\bibitem{roo} M.~Roos, ``Test of Einstein Locality'',
              HU-TFT-80-5 (revised), Nov. 1980.

\end{thebibliography}
\end{document}